# Moving Weakly Relativistic Electromagnetic Solitons in Laser-Plasmas

Lj. Hadžievski[1], A. Mančić[2] and M.M. Škorić[1]

[1] *Department of Physics, Faculty of Sciences and Mathematics, University of Niš,*
*P.O. Box 224, 18001 Niš, Serbia and Montenegro*
[2]*Vinča Institute of Nuclear Sciences, P.O.Box 522, 11001 Belgrade, Serbia and Montenegro*

**Abstract.** A case of moving one-dimensional electromagnetic (EM) solitons formed in a relativistic interaction of a linearly polarized laser light with underdense cold plasma is investigated. The relativistic Lorentz force in an intense laser light pushes electrons into longitudinal motion generating coupled longitudal-transverse wave modes. In a weakly relativistic approximation these modes are well described by a dynamical equation of the generalized nonlinear Schrödinger type, with two additional nonlocal terms [1]. An original analytical solution for a moving EM soliton case is here calculated in an implicit form. The soliton motion down-shifts the soliton eigen-frequency while decreases its amplitude. An influence of the soliton velocity on stability properties is analytically predicted.

## INTRODUCTION

Relativistic electromagnetic (EM) solitons in laser driven plasmas were analytically predicted and found by PIC (particle-in-cell) simulations [2]-[8]. It has been estimated that, for ultra-short laser pulses, up to 40% of the laser energy can be trapped by relativistic solitons, creating a significant channel for laser beam energy conversion.

Relativistic EM solitons are localized structures self-trapped by a locally modified plasma refractive index via relativistic electron mass increase and an electron density drop due to the ponderomotive force of an intense laser light [2]-[4]. A train of relativistic EM solitons is typically found to form behind the intense laser pulse front.

In this paper, we treat a case of a linearly polarized laser light. In laser-plasma interactions, relativistic Lorentz force sets electrons into motion, generating coupled longitudinal-transverse wave modes. These modes in the framework of a weakly relativistic cold plasma approximation in one-dimension, can be well described by a single dynamical equation of the generalized nonlinear Schrödinger type [1], with two extra nonlocal terms. A new analytical solution for the *moving* EM soliton case is calculated in the implicit form and the soliton motion effect on its self-frequency and amplitude is outlined. Moreover, the influence of the soliton velocity on the EM solition stability is discussed. These results are compared with the one for a standing (non-moving) relativistic EM soliton case, obtained by some of these authors [1]. Finally, numerical simulations of the model dynamical equation were performed in order to check the good agreement with our analytical results.

# DYNAMICAL EQUATIONS

We consider a long intense laser pulse propagating through a cold collisionless plasma and start with the fully nonlinear relativistic one-dimensional model for the EM wave equation, the continuity equation and the electron momentum equation, for a cold plasma with fixed ions. These equations, in the Coulomb gauge, read:

$$\left(\frac{\partial^2}{\partial t^2} - c^2 \frac{\partial^2}{\partial x^2}\right) a = -\frac{\omega_p^2}{n_0} \frac{n}{\gamma} a \tag{1}$$

$$\frac{\partial n}{\partial t} + \frac{\partial}{\partial x}\left(\frac{np}{m\gamma}\right) = 0 \tag{2}$$

$$\frac{\partial p}{\partial t} = -eE_{\parallel} - mc^2 \frac{\partial \gamma}{\partial x} \tag{3}$$

where $a = eA/mc^2$ is normalized vector potential in the $y$ direction, $n$ is the electron density, $p$ is the electron momentum in the $x$ direction, $\gamma = (1 + a^2 + p^2/m^2c^2)^{1/2}$, $E_{\parallel}$ is the longitudinal electric field, $n_0$ is the unperturbated electron density, and $\omega_p = (4\pi e^2 n_0/m)^{1/2}$ is the background electron plasma frequency.

In a weakly relativistic limit for $|a| \ll 1$ and $|\delta n| \ll 1$, the wave equation for the vector potential envelope $A$ is obtained, as (details are given in [1]):

$$i\frac{\partial A}{\partial t} + \frac{1}{2} A_{xx} + \frac{3}{16} |A|^2 A - \frac{1}{8}(|A|^2)_{xx} A + \frac{1}{24}(A^2)_{xx} A^* = 0 \tag{4}$$

Equation (4) has a form of a generalized nonlinear Schrödinger (GNLS) equation with two extra nonlocal (derivate) nonlinear terms. We can readily derive two conserved quantities: photon number $P$ and Hamiltonian $H$:

$$P = \int |A|^2 \, dx, \qquad H = \frac{1}{2} \int \left\{|A_{xx}|^2 - \frac{3}{16}|A|^4 - \frac{1}{8}[(|A|^2)_x]^2 - \frac{1}{6}|A|^2 |A_x|^2\right\} dx \tag{5}$$

Now, we look for a localized solution of (4) in a form of a moving soliton:

$$A = \rho(u) \exp[i\theta(u) + i\lambda^2 t], \tag{6}$$

where, $u = x - vt$, and $v$ is the soliton velocity. Under the boundary conditions $\rho(u), \rho(u)_u, \rho(u)_{uu} \to 0$ when $u \to \pm\infty$, the first integration of (4) gives:

$$\theta_u(u) = v, \qquad (\rho_u)^2 = \frac{2\lambda^2 - v^2 - \left(\frac{3}{16} - \frac{v^2}{6}\right)\rho^2}{1 - \frac{\rho^2}{3}} \rho^2 \tag{7}$$

Additional integration of (7) yields a moving soliton solution in an implicit form

$$\pm u = \frac{1}{2\sqrt{2\lambda^2 - v^2}} \ln \left| \frac{\sqrt{1 - \frac{\rho^2}{\rho_0^2}} + \sqrt{1 - \frac{\rho^2}{3}}}{\sqrt{1 - \frac{\rho^2}{\rho_0^2}} - \sqrt{1 - \frac{\rho^2}{3}}} \right| - \frac{1}{\sqrt{\frac{9}{16} - \frac{v^2}{2}}} \ln \frac{\sqrt{1 - \frac{\rho^2}{3}} + \sqrt{\frac{\rho_0^2}{3} - \frac{\rho^2}{3}}}{\sqrt{1 - \frac{\rho_0^2}{3}}}, \tag{8}$$

where $\rho_0^2 = \frac{2\lambda^2 - v^2}{3/16 - v^2/6}$ is the maximum amplitude of the linearly polarized moving

EM soliton with the eigen-frequency $\Lambda = \lambda^2 - v^2$. With the soliton velocity $v$ put to zero the above expression readily coincides with the standing soliton solution of the GNLS equation (4), given in [1]. Furthermore, with the ansatz (6), explicit contribution of the velocity dependent - "kinetic" terms in the Hamiltonian (7) is singled out, by

$$H = 1/2 \int \left( \rho_{uu}^2 - 2v^2 \rho \rho_{uu} - \frac{2}{3}\rho^2 \rho_u^2 + 4v^2 \rho_u^2 - \left(\frac{3}{16} + \frac{v^2}{6}\right)\rho^4 + v^4 \rho^2 \right) du . \quad (9)$$

Actually, the expression (8) for a moving EM soliton, predicts a two-parameter family of existing solutions of eq.(4). To illustrate, with the Photon number-P as the constant parameter, in figure 1, we plot stationary solutions as a function of the soliton amplitude-$\rho_0$ ($\lambda$) and its velocity- v. For P- photon energy conserved, it appears that with an increasing velocity the maximum amplitude decreases while the soliton profile broadens; up to a point when the EM localization criteria is lost. These results could be directly applicable, e.g., for a case of a slow adiabatic acceleration of isolated EM solitons in a non-uniform electron plasma.

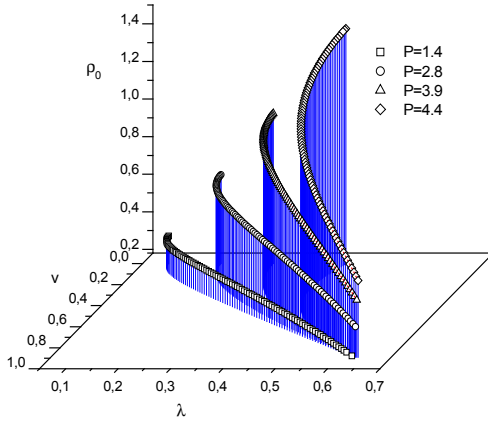
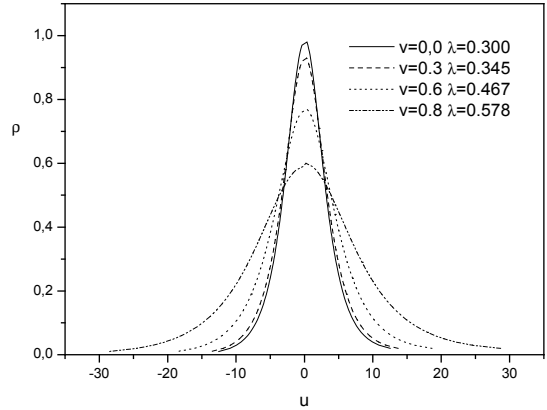

**FIGURE 1.** Soliton solutions versus the soliton amplitude $\rho_0$ ($\lambda$) and its velocity –V, for P=const

**FIGURE 2.** Broadening of the soliton profile.

## SOLITON STABILITY

In order to check the stability of the moving soliton, we use the renown Vakhitov-Kolokolov stability criterion [9,1]. According to this criterion, solitons are stable (with respect to longitudinal perturbations) if:

$$\frac{dP_o}{d\lambda^2} > 0, \quad (10)$$

where, $P_0$ is the soliton photon number defined by (5). The function $P_0(\lambda)$ can be analytically calculated for the soliton solution given by (8), as:

$$P_0(\lambda,v) = \frac{1}{\sqrt{3/16 - v^2/6}} \left[ \rho_0 + \frac{\sqrt{3}}{2}\left(1 - \rho_0^2/3\right)\ln\frac{1 + \rho_0/\sqrt{3}}{|1 - \rho_0/\sqrt{3}|} \right]. \tag{11}$$

When the soliton velocity is equal to zero above result (11) agrees with the standing soliton solution, obtained earlier in [1]. According to the condition (10), moving EM solitons turn out to be stable in the region $\lambda < \lambda_S$ (fig. 3). We conclude that small amplitude linearly polarized moving solitons within the weakly relativistic model are stable.

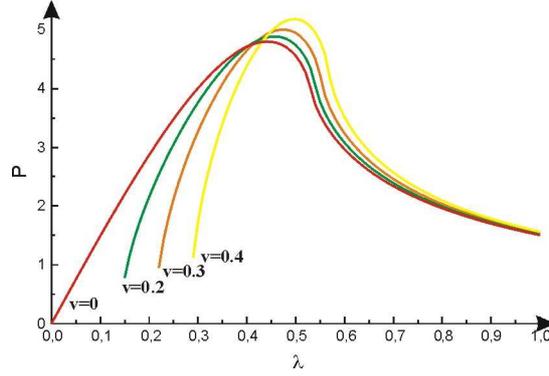

**FIGURE 3.** Photon number $P_0(\lambda)$ variation for different soliton velocities.

Comparing the Photon number $P_0(\lambda)$ variation for the moving soliton with the stationary one we see the influence of the soliton velocity on the stability region. As the soliton velocity increases, stability region are shifting toward greater values of $\lambda$.

## SIMULATION RESULTS

In order to check our analytical results and our prediction concerning the moving soliton dynamics and the influence of the velocity on soliton stability, we have performed direct numerical simulation of the nonlinear model equation (4). We have used a numerical algorithm based on the split-step Fourier method[10], which was originally developed for the NLS equation.

Numerical results prove that the initially launched moving solitons (8) with the soliton parameters inside the stability region remain stable. The evolution of the initially launched soliton with amplitude $A_0=0.4$ and photon number $P=1.8$ inside the stability region, for different soliton velocities is shown in Fig.4. The influence of the soliton velocity on its dynamics is shown in Fig. 5. Comparing the evolution of the initially launched standing soliton ($v=0$) with amplitude $A_0=1.63$ outside the stability region ($\lambda = 0.5$) (a), and the evolution of the moving soliton with the same $\lambda$ but with the velocity $v=0.7$ (b), we can see that soliton velocity acts as a stabilizing factor on its dynamics. As an illustration of an unstable behavior, a moving soliton with parameters ($\lambda$, $v$) outside the stability region is shown in Fig. 5(c).

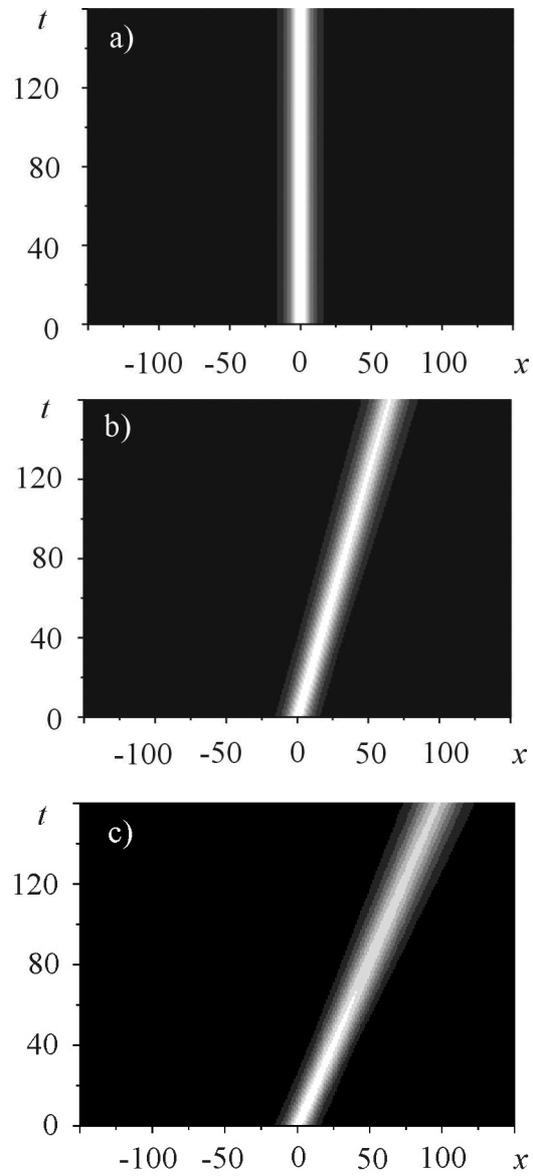

**FIGURE 4.** Spatio-temporal evolution of EM solitons, inside the stability region with amplitude $A_0$=0.4, photon number $P$=1,8 for different initial velocities: (a) $v$=0; b) $v$=0.4; and c) $v$=0.6

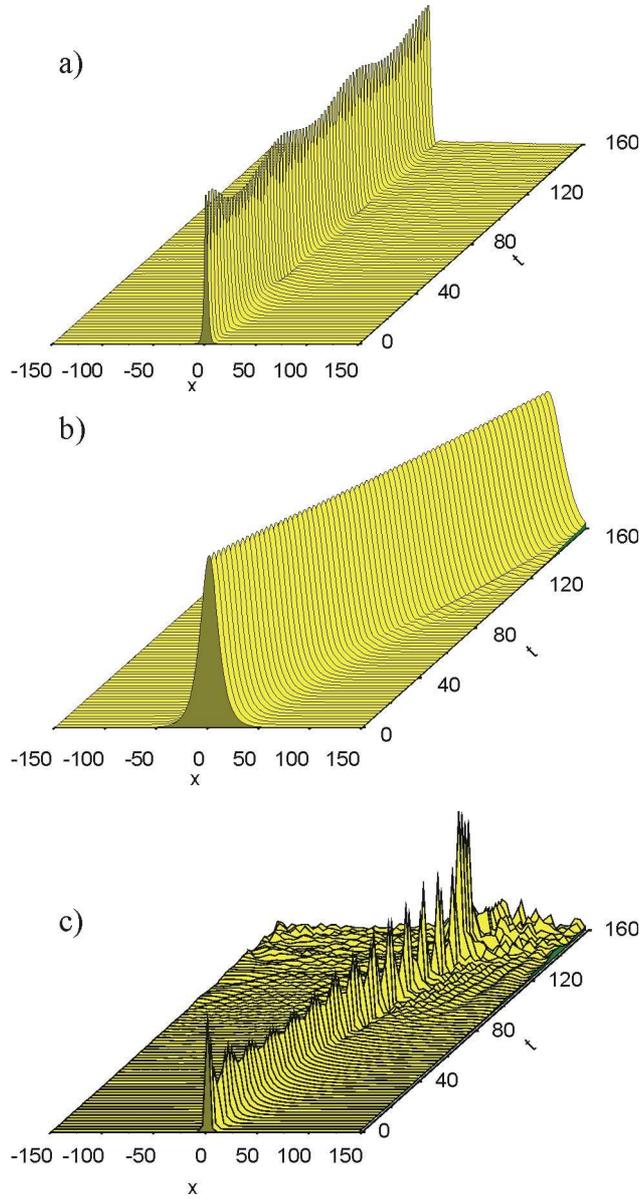

**FIGURE 5.** (a) Standing soliton in the unstable region (λ=0.5) with amplitude $A_0$=1.63; (b) Moving soliton with the same λ as in (a), $v$=0.7, $A_0$=0.3; (c) moving soliton in the unstable region, λ=0.55, $v$=0.4, $A_0$=1.66

# CONCLUSIONS

- Analytical solutions for moving linearly polarized EM solitons, described by the weakly relativistic GNLS model equation, are found for the first time.
- For an isolated EM soliton case (P= const.), a soliton velocity increase results in a reduction of the maximum amplitude and broadening of the soliton profile.
- Weakly relativistic moving EM solitons are stable; with the stability region shifting toward larger amplitudes in comparison to the standing soliton case.
- Simulations of the model GNLS equation have confirmed analytical results.

## ACKNOWLEDGMENTS

This work is supported by the Ministry of Sciences and Protection of the Environment of Republic of Serbia, Project 1964.